\documentclass[superscriptaddress,groupedaddress,nofootnoteinbib,11pt]{article}
\pdfoutput=1 
\usepackage{jheppub}

\usepackage[utf8x]{inputenc}
\usepackage{mathtools,slashed,mathrsfs}
\usepackage[caption=false]{subfig}
\usepackage{dcolumn}
\usepackage{multirow}
\usepackage{tabularx}
\usepackage{booktabs}
\usepackage{bm}
\usepackage{comment}
\usepackage{setspace}
\usepackage[dvipsnames]{xcolor}
\usepackage[normalem]{ulem} 
\usepackage{enumerate}
\usepackage{siunitx}
\graphicspath{{../../Mathematica/}}

\newcommand{\MeV}{{\, {\rm MeV}}}

\newcommand{\Tstart}{T_\textsc{r$\to$m}}
\newcommand{\Tend}{T_\textsc{m$\to$r}}

\newcommand{\frm}{f_\textsc{r$\to$m}}
\newcommand{\fmr}{f_\textsc{m$\to$r}}
\newcommand{\rtom}{\textsc{r$\to$m}}
\newcommand{\mtor}{\textsc{m$\to$r}}

\newcommand{\Mp}{M_P}

\renewcommand{\d}{\mathrm d}
\renewcommand{\H}{\mathcal H}
\newcommand{\OGW}{\Omega_\textsc{gw}}
\newcommand{\rGW}{\rho_\textsc{gw}}
\newcommand{\ts}{\tau_\star}
\newcommand{\us}{u_\star}
\newcommand{\tk}{\tau_k}
\newcommand{\Js}{J_\star}
\newcommand{\Hs}{\H_\star}
\newcommand{\fFS}{f_\textsc{fs}}
\newcommand{\neff}{N_\text{eff}}

\newcommand\bea{\begin{equation}}
\newcommand\eea{\end{equation}}
\newcommand{\new}[1]{#1}

\begin{document}

\title{Causal gravitational waves as a probe of free streaming particles and the expansion of the Universe}

\date{\today}
\author[a]{Anson Hook,}
\author[a,b]{Gustavo Marques-Tavares,}
\author[c]{and Davide Racco}

\affiliation[a]{Maryland Center for Fundamental Physics, University of Maryland, College Park, MD 20742, U.S.A.}
\affiliation[b]{Department of Physics and Astronomy, Johns Hopkins University, Baltimore, MD 21218, U.S.A.}
\affiliation[c]{Perimeter Institute for Theoretical Physics, 31 Caroline St.~N., Waterloo, Ontario N2L 2Y5, Canada}
\emailAdd{hook@umd.edu}
\emailAdd{gusmt@umd.edu}
\emailAdd{dracco@perimeterinstitute.ca}

\abstract{
The low frequency part of the gravitational wave spectrum generated by local physics, such as a phase transition or parametric resonance, is largely fixed by causality, offering a clean window into the early Universe. In this work, this low frequency end of the spectrum is analyzed with an emphasis on a physical understanding, such as the suppressed production of gravitational waves due to the excitation of an over-damped harmonic oscillator and their enhancement due to being frozen out while outside the horizon.
Due to the difference between sub-horizon and super-horizon physics, it is inevitable that there will be a distinct spectral feature that could allow for the direct measurement of the conformal Hubble rate at which the phase transition occurred.
As an example, free-streaming particles (such as the gravity waves themselves) present during the phase transition affect the production of super-horizon modes. 
This leads to a steeper decrease in the spectrum at low frequencies as compared to the well-known causal $k^3$ super-horizon scaling of stochastic gravity waves. If a sizable fraction of the energy density is in free-streaming particles, they even lead to the appearance of oscillatory features in the spectrum.
If the universe was not radiation dominated when the waves were generated, a similar feature also occurs at the transition between sub-horizon to super-horizon causality.
These features are used to show surprising consequences, such as the fact that a period of matter domination following the production of gravity waves actually increases their power spectrum at low frequencies.
}

\maketitle

\section{Introduction}

With the first direct detection of Gravitational Waves (GWs) \cite{Abbott:2016blz}, we have gained access to a new probe of the universe.  
One especially exciting aspect of GWs is that, because they are so weakly interacting, they can carry information about the early universe in a very clean and unpolluted form.
The future of GWs is bright as there are many proposed future detectors such as LISA~\cite{Audley:2017drz}, BBO~\cite{Harry:2006fi}, MAGIS~\cite{Graham:2017pmn} and DECIGO~\cite{Kawamura:2011zz}.
Using stochastic GW background searches, these new GW detectors can probe the universe at a time $10^{28}$ times earlier than our current best direct probe, the Cosmic Microwave Background (CMB). 

There are many exciting possibilities that stochastic GW backgrounds will allows us to explore.
Stochastic GW backgrounds can be produced by a myriad of early universe phenomena 
such as inflation~%
\cite{Grishchuk:1974ny,Starobinsky:1979ty,Rubakov:1982df,Guzzetti:2016mkm}%
, reheating/preheating~%
\cite{Khlebnikov:1997di,Easther:2006gt,Easther:2006vd,GarciaBellido:2007dg,GarciaBellido:2007af,Dufaux:2007pt}%
, phase transitions~%
\cite{Caprini:2019egz} %
and topological defects~%
\cite{Caprini:2018mtu,Christensen:2018iqi} %
such as cosmic strings or domain walls, or by scalar perturbations at second order in perturbation theory~%
\cite{Acquaviva:2002ud,Mollerach:2003nq,Baumann:2007zm,Espinosa:2018eve,Kohri:2018awv}.
The detection of such a background would teach us details about which of these phenomena were active and what were the relevant parameters governing their dynamics.

Aside from learning about the actual source of GWs, their discovery also teaches us about the dynamical evolution of the early universe.
There exists a vast literature discussing these effects and their detectability in various scenarios~\cite{%
Seto:2003kc,
Boyle:2005se,
Watanabe:2006qe,
Boyle:2007zx,
Jinno:2012xb,
Caprini:2015zlo,
Geller:2018mwu,
Saikawa:2018rcs,
Cui:2018rwi,
Caldwell:2018giq,
DEramo:2019tit,
Bernal:2019lpc,
Figueroa:2019paj,
Auclair:2019wcv,
Chang:2019mza,
Hajkarim:2019nbx,
Caprini:2019egz,
Gouttenoire:2019kij,
Gouttenoire:2019rtn,
Domenech:2019quo,
Blasi:2020wpy,
Domenech:2020kqm%
} as well as reviews on the subject~\cite{Caprini:2018mtu,Allahverdi:2020bys}.
Intuition for this dependence can be most easily developed focusing on a situation that has been largely overlooked, the low frequency spectrum of a stochastic GW background produced by a causal source. This is the case for GWs generated by a phase transition or parametric resonance for example.  
The low frequency modes that correspond to 
length and time scales much longer than the source duration are largely independent of the details governing the generation of GWs and are instead fixed by causality.
The largest region in causal contact with itself at the time GWs were generated has a size at most of the order of the Hubble radius, but is generically smaller. 
These low frequency modes can only be affected after entering horizon, making the spectrum sensitive to the change of the Hubble rate as a function of time, or in other words, to the expansion history of the universe.

In this paper we consider the low frequency part of the power spectrum of a stochastic background of GWs.
These modes are typically referred to as the causal part of the spectrum in order to emphasize the effect of causality, and under standard assumptions have a $k^3$ scaling~\cite{Caprini:2009fx}. Nonetheless, as we will see later, there are effects other than causality that determine their dynamics. 
We will refer to these modes as the ``causality-limited'' part of the spectrum.
The shape of the spectrum for modes which are causality-limited but were sub-horizon at generation (sub-horizon causality) is completely independent from the evolution of the universe%
\footnote{They are however sensitive to the total redshift factor between then and now.}.
Causality arguments affect modes which were super-horizon at generation in a two-fold way.
The first factor is that the production of these modes is highly suppressed.  
The reason for this is that they have frequencies $\omega \ll H$ ($H$ being the Hubble rate) at the time of production, and therefore behave like an over-damped harmonic oscillator, making such modes very hard to excite.  
As a result, this effect decreases the power at low frequencies.  
The second factor is that once excited, their amplitude stays constant while they are super-horizon, rather than decreasing, leading to an overall increase in the power at low frequencies.  
The competition between these two effects determines the shape of the low frequency tail of GWs.

More mathematically, the physical intuition that we will develop is that the initial conditions for causality-limited modes immediately after generation are $h_{ij} = 0$ while $\dot h_{ij} \ne 0$, in much the same way that the initial condition of a hammer hitting a string creates no displacement but a large velocity.
If the GW is sub-horizon, $\omega \gg H$, the mode obtains an amplitude $h \sim \dot h/\omega$.  
If the mode is super-horizon, friction is important and the mode obtains a smaller than expected amplitude $h \sim \dot h/H$.  
After this initial displacement, over-damped modes are frozen in place until Hubble crossing at which point they become the standard under-damped modes seen in GW detectors.

There are several interesting observations that result from an analysis along the previous lines.  
The first is that generically there is a difference between sub-horizon and super-horizon causality.  
As a result, there will generically be a feature at horizon crossing.  
Finding such a feature would give a model-independent direct measurement of the conformal Hubble rate at the time when GWs were created.
Upon specifying an expansion history, it is possible to deduce the Hubble rate, and hence the temperature, at which the GWs were generated.
We will discuss two scenarios in which GW modes that are super-horizon or sub-horizon (but at a wavelength larger than the correlation length of the source) at the time of generation behave differently.

In the first, we study how super-horizon GWs depend on the equation of state parameter $w$.  
We show that while sub-horizon causality forces the spectrum of GWs to fall as $k^3$, that super-horizon modes fall as $k^{(1+15 w)/(1+3 w)}$.  Thus, as long as $w \ne 1/3$, the scaling of GWs changes between modes that are sub-horizon at the time of generation versus super-horizon.
The change in behavior between sub-horizon modes and super-horizon modes during matter domination (MD) was previously studied in Ref.~\cite{Barenboim:2016mjm} and followed up in Refs.~\cite{Cai:2019cdl,Guo:2020grp,Ellis:2020nnr}. 
We improve upon these results by developing simple physical intuition for these effects and by generalizing them to arbitrary equations of state.

For the second scenario we look at the effect of free-streaming particles on the production and propagation of GWs.  
This was originally studied in Ref.~\cite{Weinberg:2003ur} for the case where some species of particles started free-streaming after GWs were produced, and it showed that this leads to a constant suppression in the amplitude as modes entered the horizon.  
We focus on the contrasting case where free-streaming particles were already present during the creation of the GWs.  We find that sub-horizon modes have the standard $k^3$ scaling.  Super-horizon modes have a scale dependent suppression, and for a sufficiently large fraction of free-streaming particles there is a a surprising oscillatory feature on top of a suppressed scaling of $k^4$.  
These effects come about because in this scenario super-horizon GWs are not frozen out but instead slowly roll in the potential of the free-streaming particles, suppressing the power in GWs.  This effect is always present, even if at a small level, as the high frequency part of the GW spectrum itself acts as free-streaming particles.

Another interesting result we find is that an intermediate period of matter domination after the GW generation actually increases the power at low frequencies.  
There are well motivated scenarios where various effects due to reheating generate GWs~\cite{Khlebnikov:1997di,Easther:2006gt}. 
In the typical scenario where reheating occurs at high scales, only the causal part of the GW spectrum is accessible as the peak frequencies are too high.  
An additional feature of many of these models is that they generically imply matter domination after the production of GWs.
This effect was previously considered to be undesirable as it dilutes the peak power of the GW. However, for the low frequencies that are experimentally visible, the effect of a period of matter domination is to increase the visibility of these scenarios, making it a desirable feature.

In Sec.~\ref{Sec: spectrum}, we discuss the physical intuition behind causality-limited GWs and compare our physically motivated approximations against exact results.
In Sec.~\ref{Sec: free}, we study the effects of free streaming particles on causal GWs.
In Sec.~\ref{Sec: matter}, we study how the frequency spectrum of gravitation waves changes as the universe transitions between matter, radiation, and sub-horizon modes.
Finally, we conclude in Sec.~\ref{Sec: conclusion}.

\section{The spectrum of causality-limited gravity waves} \label{Sec: spectrum}

\subsection{The physical intuition} 
\label{Sec: intuition}

Throughout this work we will be exploring ``causality-limited" GWs, that we define as follows.  
Consider a source that is active for a short amount of time $1/\beta$.  
The causality-limited part of the GW spectrum consists of the waves whose period and wavelength are much longer than the source's temporal and spatial correlations respectively.  In other words, $\lambda \gg 1/\beta$, where $\lambda$ is the wavelength and $1/\beta$ is the duration of the process generating the GWs.  
The standard example of such causality-limited GWs is the low frequency part of the GW spectrum generated by a cosmological phase transition.

Using conformal time $\tau$ and conformal Hubble rate $\H=a'/a$, the equation of motion for a comoving mode $k$ of the graviton $h_{ij}$ is (we follow a notation close to the one of \cite{Caprini:2009fx})
\bea
\label{Eq: eom}
\partial^2_\tau h_{ij} + 2 \H \partial_\tau h_{ij} + k^2 h_{ij} = a^2 \frac{32 \pi G \rho}{3} \Pi_{ij} \equiv J_{ij} \,,
\eea
\new{where $\Pi_{ij}$ is the dimensionless anisotropic stress of the sector generating the GWs, $\rho$ is the energy density of that sector, and the normalization of $\Pi_{ij}$ follows that of \cite{Caprini:2009fx} with a prefactor of $32\pi G/3$.}
We will project both $h_{ij}$ and $J_{ij}$ onto the two independent $(+,\times)$ polarisations, and assume that the respective amplitudes $J$ and $h$ of the two polarisations are equal.
We will also assume that the GWs are produced on a time scale fast compared to Hubble so that we can approximate $\H$ as a constant over its production and the time over which $\Pi$ is non-zero to be small. 
The time at which the phase transition occurs will be denoted by $\ts$.  

The solution to Eq.~\eqref{Eq: eom}\new{, assuming $h(\tau)=0$ for $\tau < \tau_\star$,} can be found using Green's functions to be
\bea
\label{Eq: intermediate}
h(k,\tau) = \int \mathrm d\tau' \frac{e^{-\H (\tau - \tau')}}{\sqrt{k^2 - \H^2}} \sin \left( (\tau-\tau') \sqrt{k^2 - \H^2} \right) J(k, \tau') \, .
\eea
We will be interested in the initial conditions right afterwards $\tau' = \ts + \epsilon$ assuming that $J$ occurs fast so that we can take $J(k,\tau') = \Js(k) \delta (\tau' - \ts)$, and we will drop the $k$ dependence from here on.
\new{The approximation of the source as a Dirac delta simplifies the analysis and holds for modes whose period is much longer than the duration of the source, and thus cannot resolve details of the time dependence. The range of modes for which the approximation holds will vary depending on the exact dynamics of what sources the gravitational wave.}
Using this approximation and Eq.~\eqref{Eq: intermediate}, we find that
\bea
h (k,\ts + \epsilon) = 0 \,, \qquad 
\partial_\tau h (k, \ts + \epsilon) = \Js \,.
\eea
Afterwards, we have the second order differential equation and initial conditions
\begin{equation}
\begin{gathered}
\label{Eq: master}
\partial^2_\tau h + 2 \H \partial_\tau h + k^2 h = 0 \,,\\
h (k,\ts) = 0 \,, \quad \, \partial_\tau h (k, \ts) = \Js \,. 
\end{gathered}
\end{equation}

The observable relevant for GW detectors such as LISA is $\d \, \OGW/\d\log k$.  
This is obtained using
\begin{equation}
\rGW(\mathbf x,\tau) = \sum_{r,s=+,\times} 
  \frac{1}{32 \pi G a^2} \left \langle 
  h^{\prime\, r}_{ij}(\mathbf x,\tau) 
  h^{\prime\, s}_{ij}(\mathbf x,\tau) \right\rangle \,,
\end{equation}
\new{which holds only when the relevant modes have entered the horizon at late times and started oscillating.}
We will also take
\begin{equation}
\left\langle h(k,\tau) h(k',\tau) \right\rangle \equiv \left ( 2 \pi \right )^3 \delta^3 (k-k') P_h(k,\tau) \,,
\end{equation}
with $P_h(k,\tau)$ the dimensionful power spectrum of GWs. 
Fourier transforming gives
\bea
\frac{\d \OGW}{\d \log k} = \frac{1}{\rho_c} \frac{\d \rGW}{\d \log k} = \frac{1}{\rho_c} \frac{k^5 P_h(k,\tau)}{2 \left ( 2 \pi \right )^3 a^2 G} \, ,
\eea
where $\rho_c = 3 H^2/(8 \pi G)$ is the critical energy density. 
In the second equality we use the simplification valid at late times that $h^\prime=k\, h$, and where an additional $k^3$ comes from the phase space factor. 
We can then obtain the $k$ scaling of $\d \, \OGW/\d\log k$ by taking $k^5$ and multiplying by the $k$ scaling of $h^2$. 

The observable of interest depends on $h(k, \tau)$, which comes from solving Eq.~\eqref{Eq: master}. From that we can see that there are two contributions to the $k$ scaling of $\langle h h \rangle$, the $k$ dependence of the source $\Js (k)$ and propagation effects. 
Here is where causality plays a central role: as long as we are interested in wavelengths much longer than the typical spatial correlation of the sources, the Fourier transformed source $\Js \approx \text{constant}$ (see e.g.\ \cite{Caprini:2009fx}), and hence the $k$ dependence coming from the source is trivial for long wavelength modes.%
\footnote{More explicitly, if for separations $x$ larger than a length scale $\lambda$ the two-point function $P_\Pi(x)$ of the source vanishes, then its Fourier transform 
$P_\Pi(k)
 = \int \d^3\mathbf x e^{i\mathbf k\cdot \mathbf x} P_\Pi(x) 
 = \int_0^\lambda \d x \, 4\pi x\, \frac{\sin(kx)}{k} P_\Pi(x) \approx \int_0^\lambda \d x \, 4\pi x^2\, P_\Pi(x)$ is a constant independent of $k$ when $k \ll \lambda^{-1}$.}
Assuming correlation lengths for the source, $k_\text{source}^{-1}$, that are small compared to the horizon size, we can separate the solutions of the causality-limited modes into two regimes, $k_\text{source} \gg k \gg \Hs$ (sub-horizon) and $k \ll \Hs$ (super-horizon).
Note that in the case of the phase transitions, the existence of a causal sub-horizon regime depends on whether all the sources of GWs (including in particular the sound wave contribution) have a short duration compared to the Hubble time (see e.g.~\cite{Jinno:2017fby,Konstandin:2017sat,Cutting:2020nla,Jinno:2020eqg} for recent studies on this topic).

\paragraph{Sub-horizon regime $k \gg \Hs$ \, :}
These modes are under-damped by definition.  
Thus, the standard approximation is that of a frictionless solution rescaled by $a(\ts)/a(\tau)$ to take care of Hubble friction as can be seen by using the WKB approximation.  
This is easily shown to be
\bea
\label{Eq: sub-horizon}
h \approx \frac{a(\ts)}{a(\tau)} \frac{\Js}{k} \sin  k \left (\tau - \ts \right ) 
\,, \qquad (k\gg \Hs)\,.
\eea
From this, we immediately see that sub-horizon causality automatically gives 
\bea
\frac{\d \OGW}{\d \log k} \propto k^3
\,, \qquad (k\gg \Hs)
\eea
regardless of what the equation of state of the universe is.

\paragraph{Super-horizon regime $k \ll \Hs$ \, :}
There are two competing factors controlling the dynamics of the $k \ll \Hs$ modes.  The first is that one is attempting to excite an over-damped harmonic oscillator.  This effect suppresses the power in these modes.  The second is that super-horizon modes are frozen in place until they enter the horizon, increasing the relative power in these modes.  
The competition between these two effects is what controls the behavior of $k \ll \Hs$ modes.

Very quickly (roughly after a single $e$-fold) the initial conditions
\bea
h (k,\ts) = 0 \qquad \partial_\tau h (k, \ts) = \Js 
\eea
evolve into 
\bea
h (k,\ts) \sim \frac{\Js}{\Hs}\,, \qquad 
\partial_\tau h (k, \ts) \approx 0 
\label{eq: initial intuition}
\eea
due to Hubble friction, and remain frozen while the mode is super-horizon.  
The $\mathcal{O}(1)$ number out front depends on the exact equation of state of the universe.  
This scaling can be confirmed by solving $\partial^2_\tau h_{ij} + 2 \Hs \partial_\tau h_{ij} = 0$ with a non-zero initial velocity.
Notice that the amplitude of the oscillation is suppressed compared to what one would expect based on the amplitudes for sub-horizon modes: $\sim \Js/k$.  
This suppression is the effect of attempting to excite an over-damped harmonic oscillator.

At this point, the mode is frozen out and its amplitude is constant horizon entry, which occurs at a conformal time $\tk$ given by
\bea
\H(\tk) \approx k.
\eea
After the mode enters the horizon, the GW can be approximated as completely under-damped and redshifts away as $a(\tk)/a(\tau)$ giving
\bea
\label{Eq: super-horizon}
h \approx  \frac{a(\tk)}{a(\tau)} \frac{\Js}{\Hs} \sin  k \tau 
\,, \qquad (k\ll \Hs)\,.
\eea
We have been cavalier about the phase information just putting in $\sin k \tau$, as the phase is unimportant for computing the power due to averaging over many oscillations.

\subsection{Radiation domination}

It is well known that in radiation domination (RD) there is a simple solution for the GWs. 
During RD, $a \sim \tau$ and $\H = a'/a = 1/\tau$.  
We can solve Eq.~\eqref{Eq: master} exactly in this case and find
\bea
h = \frac{\Js \ts}{k \tau} \sin k \left (\tau - \ts \right ) = \frac{a(\ts)}{a(\tau)} \frac{\Js}{k} \sin k \left (\tau - \ts \right) \,.
\eea
Thus, the full solution is given by the solution in a friction-less universe times a red-shifting factor of $a(\ts)/a(\tau)$.

We now show that the physical intuition given in Sec.~\ref{Sec: intuition} replicates this exact solution.  
The sub-horizon modes given in Eq.~\eqref{Eq: sub-horizon} have the same scalings as the exact solution.
The super-horizon modes given in Eq.~\eqref{Eq: super-horizon} simplify using $k = \H(\tk) = \Hs \, a(\ts)/a(\tk)$ to give
\bea
h \approx \frac{a(\tk)}{a(\tau)} \frac{\Js}{\Hs} \sin k \tau \sim \frac{a(\ts)}{a(\tau)} \frac{\Js}{k} \sin k \tau
\eea
which is again the same scaling as the exact solution, confirming our physical intuition.

From these results we can calculate how $\d \OGW/\d\log k$ scales with $k$:
\bea
h(k) \sim \frac{1}{k} \quad \Rightarrow \quad
\frac{\d \OGW}{\d \log k} \sim k^5 h(k)^2 \sim k^3 
\,, \qquad (k\ll \Hs)\,.
\eea
This reproduces the famous fact that causality in a RD universe leads to the spectrum of GWs falling off as $k^3$ in the small $k$ limit.

\subsection{General equation of state}
We now repeat our exercise for a general equation of state $p = w \rho$.
In this case, we have 
\bea
a \propto \tau^{n} \,, \qquad \H =  \frac n\tau \,,
\eea
with $n =2/(1+3w)$, so that for radiation (matter) domination we have $n=1$ ($n=2$). In this case the exact solution of Eq.~\eqref{Eq: master} is
\begin{equation}
\label{Eq: exact solution}
h(k,\tau) = \Js k \ts^2 \left ( \frac{\ts}{\tau} \right )^{n-1} 
\Big[j_{n-1} (k \ts) y_{n-1} (k\tau) - j_{n-1} (k\tau) y_{n-1} (k \ts)\Big] \, ,
\end{equation}
where $j_n(x)$ and $y_n(x)$ are spherical Bessel functions of first and second kind respectively.
While this is the exact solution, it is useful to consider two limits.  
The first is the limit of sub-horizon modes:
\bea
\label{eq: general-sub}
h(k,\tau) \approx \frac{a(\ts)}{a(\tau)} \frac{\Js}{k} \sin k \left(\tau - \ts \right) 
\,, \qquad (k\gg \Hs)\,.
\eea
The second is the limit of super-horizon modes at production after they enter the horizon, $\Hs \gg k \gg \H$:
\bea
h(k,\tau) \approx \frac{\Gamma\left(n-\frac{1}{2} \right) \Js \ts}{2 \sqrt{\pi}} \left( \frac{2}{k\tau} \right)^n \cos \left( k\tau - \frac{n \pi}{2} \right) 
\,, \qquad (k\ll \Hs)\,.
\eea

Again, we will show that the physical intuition given in Sec.~\ref{Sec: intuition} replicates this exact solution in the two limits. The sub-horizon solution in Eq.~\eqref{eq: general-sub} has the same parametric behavior found in Eq.~\eqref{Eq: sub-horizon}. For the super-horizon case, note that Eq.~\eqref{Eq: super-horizon} for a general equation of state can be simplified using
\bea
\H(\tk) = \Hs \, \left ( \frac{a(\ts)}{a(\tk)} \right )^{\frac 1 n} = k \, .
\eea
After some algebra, we find that
\bea
h \approx 
\frac{a(\tk)}{a(\tau)} \frac{\Js}{\Hs} \sin k \tau  \sim 
\frac{\Js \ts}{\left( k\tau \right)^{n}} \sin k \tau \, ,
\label{eq: super-horizon general}
\eea
reproducing the parametric behavior of the exact solution for the super-horizon modes.

The scaling of $\d \OGW/\d \log k$ can be directly obtained from the scaling of the solutions.  For sub-horizon modes $k \gg \Hs$
\bea
\frac{\d \OGW}{\d \log k} \sim k^5 h(k)^2 \sim k^3 
\,, \qquad (k\gg \Hs)\,.
\eea
As expected, as long as the wavelength of the mode is sufficiently larger than the distances over which the sources are spatially correlated, sub-horizon causality forces these modes to fall as $k^3$. 
For the super-horizon modes, we have
\bea
\label{Eq: superhorizonk}
\frac{\d \OGW}{\d \log k} \sim k^5 h(k)^2 \sim k^{5-2n} 
\,, \qquad (k\ll \Hs)\,.
\eea
We see that there is more power at low $k$ for any equation of state with $w < 1/3$.  In particular, for the case of MD, $n=2$, we find that
\bea
\frac{\d \OGW}{\d \log k} \sim k \qquad \text{Matter domination}.
\eea

From the discussion above we find that there is a distinct kink in the spectrum at horizon crossing as the modes transition from sub-horizon to super-horizon causality.
This kink in the spectrum occurs for all cosmologies, except for exact RD where the spectrum scales as $k^3$ for long-wavelength modes. Identification of this generic feature in a signal allows for a model independent measurement of the value of conformal Hubble at which the GWs were generated. 

\begin{figure}[h!] \centering
\includegraphics[width=.6\textwidth]{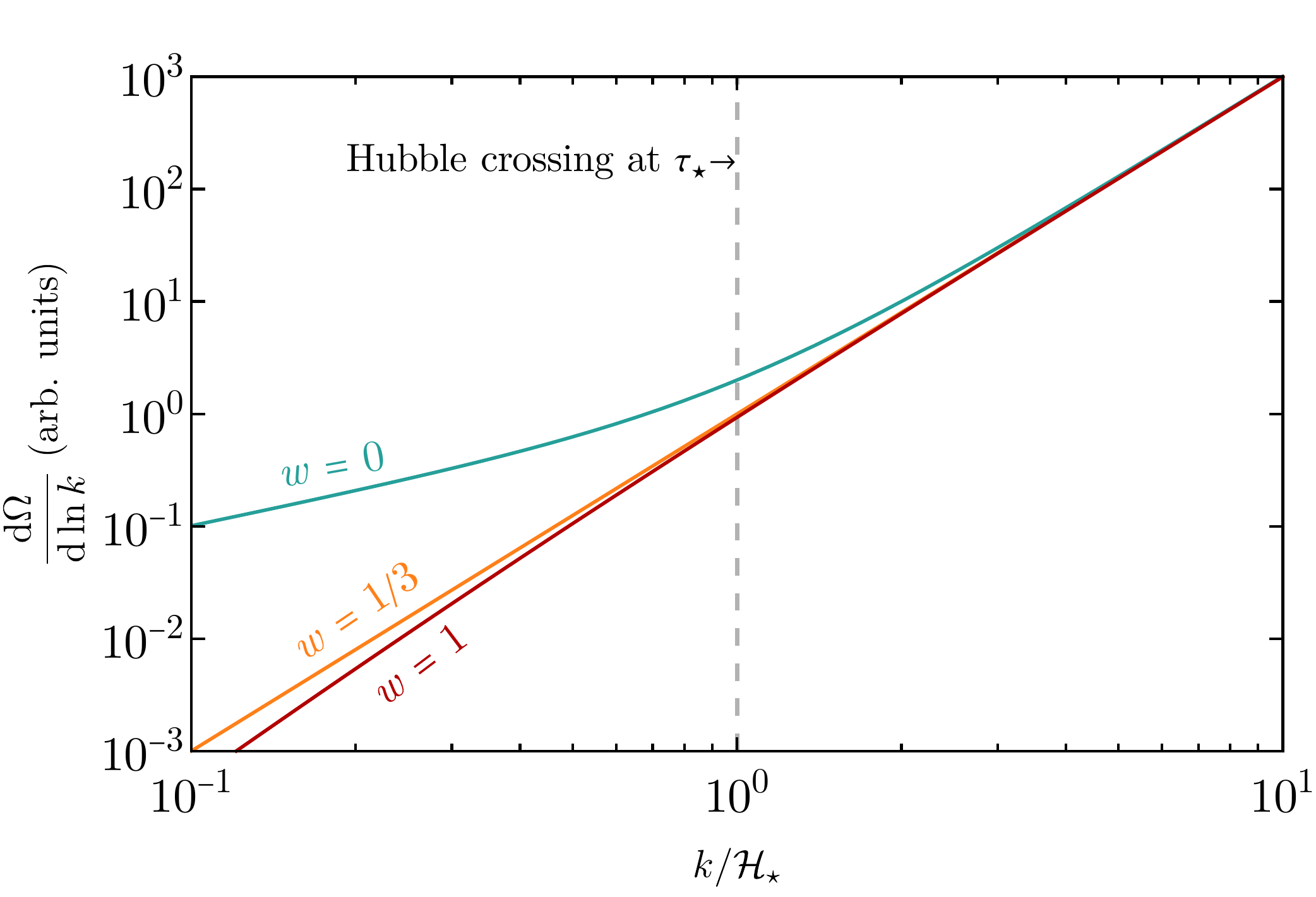}
\caption{Plot of the scaling of $\d \OGW/\d \log k$ versus $k$ for different equations of state $w$.  
The different cosmologies are all normalized so that their sub-horizon modes are of the same size in order to emphasize their super-horizon differences.}
\label{Fig: analytic}
\end{figure}

We can use the exact analytic results of Eq.~\eqref{Eq: exact solution} to obtain $\d \OGW/\d \log k$.  To do this, we take the late time limit of Eq.~\eqref{Eq: exact solution} and find
\begin{equation}
	h(k,\tau) \approx \Js \ts \left(\frac{\ts}{\tau}\right)^n \left[ j_{n-1}(k\ts) \sin\left(k\tau -\frac{n \pi}{2}\right) -y_{n-1}(k\ts) \cos\left(k\tau -\frac{n \pi}{2}\right) \right] \, .
\end{equation}
From the equation above we quickly see that
\bea
|h|^2 =  \Js^2 \ts^2 \left(\frac{\ts}{\tau}\right)^{2 n} \left(|j_{n-1}(k\ts)|^2+|y_{n-1}(k\ts)|^2 \right) \sin^2(k\tau + \phi) \, ,
\eea
where $\phi$ is a constant phase factor.
Finally, we arrive at an exact expression for the energy in GWs (after averaging over the oscillations)
\bea
\frac{\d \OGW}{\d \log k} \propto k^5 \langle h^2 \rangle \propto k^5 \left(|j_{n-1}(k\ts)|^2+|y_{n-1}(k\ts)|^2 \right)
\eea
where we have neglected unimportant proportionality constants.
In Fig.~\ref{Fig: analytic} we use these exact analytic results to show how sub-horizon scaling becomes super-horizon scaling for various equations of state.

\section{The effect of free-streaming particles} \label{Sec: free}

In this section, we show how free-streaming relativistic particles induce not only a dampening of GWs at horizon crossing but also change the shape of the causality-limited part of the GW spectrum.
We specialize to the case of a radiation-dominated universe with a fraction $\fFS = \rho_\textsc{fs}/\rho_\text{total}$ in free-streaming relativistic particles since for other equations of state the fraction $\fFS$ changes significantly with the expansion of the universe.
As was derived in Ref.~\cite{Weinberg:2003ur}, GWs are affected by the presence of particles whose free-streaming length is larger than the Hubble radius. 
In this case, the propagation of the free-streaming particles along geodesics is affected by and affects the propagation of GWs. 

It is well known that during BBN and CMB a large fraction of the energy density in radiation was free-streaming, since neutrino interactions freeze out at temperatures below approximately $2 \MeV$. Little is known about our Universe prior to this era, and therefore the existence of other relativistic free-streaming species at earlier times remains an interesting possibility in the early Universe. 
In fact, since GWs themselves are free-streaming, the high frequency part of the GW spectrum itself is an irreducible contribution to the population of free streaming particles.
We will see that if such particles were present during the generation of GWs, they would lead to striking features in the causality-limited part of the spectrum.
	
The equation of motion for GWs in this scenario is \cite{Weinberg:2003ur}
\begin{multline}
\label{Eq: weinberg}
u^2 \partial^2_u h(u) + 2 u \partial_u h(u) + u^2 h(u) = \\
-24 \fFS \int_{u_\star}^{u} \mathrm d x \left( - \frac{\sin (u-x) }{(u-x)^3}  - 3 \frac{ \cos (u-x)}{(u-x)^4} + 3 \frac{\sin (u-x)}{(u-x)^5} \right )\partial_x h (x) \,,
\end{multline}
where $u = k \tau$.
As indicated before, we are interested in causality-limited GWs.  
Thus, we are solving these equations subject to the boundary condition
\bea
h (k,\ts) = 0 \,, \qquad \partial_\tau h (k, \ts) = \Js \,. 
\eea

There are two important effects that free-streaming particles have on GWs: post-pro\-duc\-tion dynamics and horizon entry. 
The effects of free-streaming particles on horizon entry is the standard effect emphasized in the case of neutrinos. 
Note that the LHS of Eq.~\eqref{Eq: weinberg} is larger than the RHS by a factor of $u^2$ and hence the largest effect is expected to be at early times when $u = k \tau \lesssim 1$.
From this we see that sub-horizon modes are completely unaffected by free-streaming particles. Conversely, super-horizon modes are hit by a uniform frequency independent suppression when they enter the horizon, e.g. GW amplitudes are suppressed by a factor of $0.8$ when $f = 0.4$, as originally shown in Ref.~\cite{Weinberg:2003ur}.  
Amusingly, we will see that for GWs generated by phase transitions, this frequency-independent suppression is in fact a subdominant effect.

The novel effect here is that free-streaming particles also cause a suppression in the production of GWs.  
What we mean by this is that the transition of super-horizon GWs from a large velocity with a small amplitude to a large amplitude with no velocity discussed in Eq.~\eqref{eq: initial intuition} is significantly affected by the presence of free-streaming particles.
The suppressed production can be computed analytically using the techniques of the previous sections.

We first consider the limit $u=k \tau \ll 1$, where the modes are very super-horizon.  
In this limit, Eq.~\eqref{Eq: weinberg} can be simplified to 
\bea
\label{Eq: k=0}
u^2 \partial^2_u h(u) + 2 u \partial_u h(u) + u^2 h(u)= - \overline \fFS \Big ( h(u) - h(u_\star) \Big )\,,
\qquad \overline \fFS = \frac{8 \fFS}{5} .
\eea
The only difference between this equation and Eq.~\eqref{Eq: weinberg} is that we have taken the $k \tau \ll 1$ limit of the RHS, eliminating the integral. This approximation allows us to study the effect of free-streaming radiation on the evolution of super-horizon GWs. The approximation breaks down near horizon entry, however the effect of horizon entry has been studied in the past and approximately amounts to a $k$-independent reduction of the amplitude, proportional to $\fFS$ for $\fFS \ll 1$, and so it can be treated separately.  

The suppression of GW production by free-streaming particles can be directly seen from the RHS of Eq.~\eqref{Eq: k=0}, where the free-streaming particles act like a Hubble scale mass.
Thus, in contrast to the normal scenario where wave amplitudes freeze out and remain constant while outside of the horizon, the effective Hubble scale mass is constantly reducing the amplitude of the GW due to a slow roll type effect. 
Note that this effect is absent if the wave starts out with negligible velocity like in previously studied scenarios, in which cases the RHS vanishes.

Eq.~\eqref{Eq: k=0} can be solved exactly with our initial conditions to give
\begin{equation}
\label{eq: ffs-solution}
\begin{aligned}
h(\tau) &=  \frac{\Js \us^2}{k} \left [ j_\alpha(\us) y_\alpha(u) - j_\alpha(u) y_\alpha(\us) \right ] \qquad \alpha = \frac{-1 + \sqrt{1-4 \overline \fFS}}{2}  \\
&\approx - \frac{\Js \us^2}{u k} \left [ j_\alpha(\us) \cos\left( u - \frac{ \alpha \pi}{2} \right) - y_\alpha(\us) \sin\left( u - \frac{\alpha \pi}{2} \right) \right ] ,
\end{aligned}
\end{equation}
where $j_\alpha$ and $y_\alpha$ are spherical Bessel functions and we have taken the large time limit in the second line.  From this, we find that the suppression factor is simply
\begin{equation}
\label{eq: ratio}
\frac{h_\text{amplitude}(\overline \fFS)}{h_\text{amplitude}(\overline \fFS=0)} \approx \sqrt{\us^2 j_\alpha(\us)^2 + \us^2 y_\alpha(\us)^2 } \, ,
\end{equation}
which follows directly from Eq.~\eqref{eq: ffs-solution} (and holds even when $\alpha$ is imaginary).

While Eq.~\eqref{eq: ratio} is rather un-illuminating in and of itself, it can be simplified in two interesting limits $4 \overline \fFS \ll 1$ and $4 \overline \fFS > 1$.  
In the first limit, we find that
\bea
\frac{h_\text{amplitude}(\overline \fFS)}{h_\text{amplitude}(\overline \fFS=0)} \approx (k \ts)^{\overline \fFS} 
\qquad\qquad
\left(\overline \fFS \ll 1 \right) .
\eea
In the other limit, $4 \overline \fFS > 1$, we find 
\begin{equation}
\frac{h_\text{amplitude}(\overline \fFS)}{h_\text{amplitude}(\overline \fFS=0)} 
\approx \sqrt{k \ts} \sqrt{ C_1 + C_2 \sin \left (  \sqrt{4 \overline \fFS -1} \log ( k \ts )  + C_3 \right ) } ,
\label{Eq: large}
\end{equation} 
where $C_1$, $C_2$, $C_3$ are unenlightening functions of $\fFS$.
Interestingly, we find that in this limit there is an overall suppression of the amplitude that is independent $\fFS$ and that there is an additional oscillatory feature on this suppressed amplitude.  
The appearance of an oscillation as $\fFS$ increases comes from when the super-horizon modes go from being over-damped to under-damped even while super-horizon.  In this limit, the mass coming from free-streaming particles overcomes Hubble friction and induces oscillations.

To obtain the shape of the GW spectrum as a function of $k$, we numerically solve Eq.~\eqref{Eq: weinberg}. 
\begin{figure}[h!] \centering
\includegraphics[width=.75\textwidth]{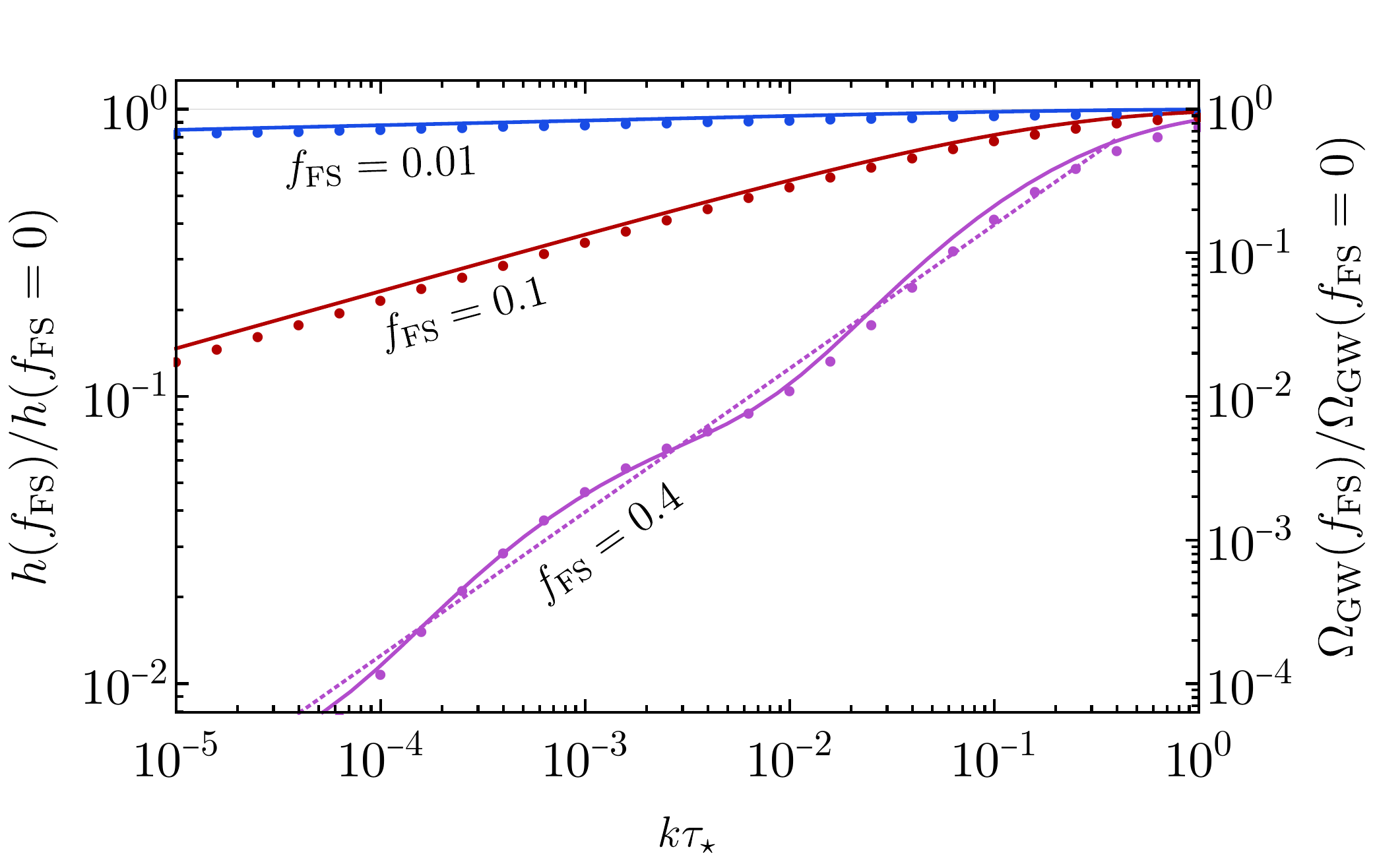}
\caption{The suppression effect of free streaming particles on the production of GWs, for free streaming fractions 0.01, 0.1 and 0.4 respectively.  
The dots are numerical data points while the solid lines are the analytical approximations.  
The dotted purple line is a $k^4$ scaling shown to highlight the oscillations taking place.}
\label{fig: suppression}
\end{figure}
The results for the suppression can be found in Fig.~\ref{fig: suppression}, which shows the numerical results as well as the analytical approximation, Eq.~\eqref{eq: ratio}.  
As can be clearly seen, the analytic results and the numerical agree very well for $\fFS = 0.01, 0.1$ and $0.4$.  This agreement holds despite the fact that the transition region, where $u =  k \tau \sim 1$, is not accurately captured by our approximation, and hence, the traditional effect of horizon entry is not entirely taken into account.

We can re-express our results for the amplitude of the GWs in terms of the observable $\d \OGW/\d \log k$.  For sub-horizon modes, we have the previous result
\bea
\frac{\d \OGW}{\d \log k} \sim k^3 
\qquad \qquad 
\left(k \gg \Hs \right). 
\eea
For super-horizon modes, we have
\bea
\frac{\d \OGW}{\d \log k} \sim k^{3 + \frac{16}{5} \fFS} 
\qquad \qquad 
\left(k \ll \Hs \, , \ \fFS \ll 1 \right)
\eea
for small free-streaming fractions and
\bea
\frac{\d \OGW}{\d \log k} \sim k^{4} \left( C_1 + C_2 \sin \left (  \sqrt{4 \overline \fFS -1} \log ( k/ \Hs )  + C_3 \right ) \right ) 
\qquad 
\left(k \ll \Hs  \, , \ \fFS \gtrsim \frac{5}{32} \right)
\eea
for large free-streaming fractions, where $C_{1,2,3}$ are constants. This shows that the presence of free-streaming radiation during the generation of GWs leads to a spectrum that decreases faster at low frequency, with the spectral index growing linearly in $\fFS$ from $3$ until it saturates at $4$ for $\fFS \geq \frac{5}{32}$.

As mentioned before, the high $k$ (sub-horizon) modes of the GWs themselves necessarily play the role of free-streaming particles.  
The derivation of Eq.~\eqref{Eq: weinberg} assumes that the free-streaming particles follow geodesics, which GWs do as well.  
Thus, the only possible difference between the high frequency GW modes and free-streaming particles is that the high frequency GW modes were produced by the phase transition itself.
Imagine that the GWs are produced at a time $\epsilon$ after the phase transition, then Eq.~\eqref{Eq: weinberg} becomes
\begin{multline}
u^2 \partial^2_u h(u) + 2 u \partial_u h(u) + u^2 h(u) = \\
-24 \fFS \int_{u_\star+k\epsilon}^{u} \mathrm d x \left( - \frac{\sin (u-x) }{(u-x)^3}  - 3 \frac{ \cos (u-x)}{(u-x)^4} + 3 \frac{\sin (u-x)}{(u-x)^5} \right )\partial_x h (x) \,.
\end{multline}
As long as $\Hs \epsilon \ll 1$, we recover Eq.~\eqref{Eq: k=0} and we see that the high frequency GW modes behave exactly like any other free-streaming particles.  

\subsection{Implications of current $N_\text{eff}$ limits}

Results from CMB and BBN measurements limit the amount of non Standard Model radiation present in the Universe at those times. These results are presented as limits on the effective number of neutrinos,
\begin{equation}
	\neff = \frac{8}{7} \left( \frac{11}{4} \right)^{4/3} \frac{ (\rho_\nu + \rho_X) }{\rho_\gamma} \, ,
\end{equation}
where $\rho_\gamma$ and $\rho_\nu$ are respectively the energy density in photons and neutrinos and $\rho_X$ represents any new contribution to the energy density in radiation. 

A combined analysis of BBN and CMB measurements leads to a bound $\Delta \neff \lesssim 0.3$ \cite{Aghanim:2018eyx}. This implies that at neutrino decoupling a new type of radiation could contribute to at most $4.4\%$ of the radiation energy. However, due to the decoupling of massive degrees of freedom, the relative temperature between the standard model plasma and any decoupled species changes as the Universe evolves. In particular, a $\Delta \neff \lesssim 0.3$ at BBN translates to $\fFS \lesssim 0.09$ for temperatures above the weak scale. 

This shows that if below the weak scale there are no modifications to $\Lambda$CDM, the maximum allowed $\fFS$ is $0.09$, and if a non-zero $\neff$ is measured in future CMB experiments, the nature of this radiation at high energies could be tested using GWs.%
\footnote{While BBN measurements alone cannot distinguish free-streaming radiation from an interacting relativistic fluid, CMB measurements are now precise enough to differentiate between them (see e.g.~\cite{Baumann:2015rya,Brust:2017nmv,Blinov:2020hmc}). Note that the CMB measurements are only sensitive to the nature of the relativistic species at low energies, $\mathcal{O}$(eV).
}
Nevertheless larger $\fFS$ are possible with simple modifications of early cosmology, such as late equilibration of the radiation with the SM with subsequent decay or through an entropy injection into the standard model plasma, which dilutes the relative energy density in the new species. The latter case would also lead to a temporary change in the equation of state that changes the GW signal as discussed in the next section.

\section{Effect of non-standard expansion histories} \label{Sec: matter}

In this section we show how the causality-limited spectrum of GWs can be used to probe the expansion rate of the Universe after the GWs were generated. There are two main effects of a non-standard expansion history in the spectrum of GWs: an overall re-scaling of the spectrum due to the change in expansion history; and a change in the power-spectrum of causality-limited modes that enter the horizon during an era of non-radiation domination. The first feature has been widely discussed in the literature, while the second has been studied in the case of tensor modes for inflation \cite{Saikawa:2018rcs,DEramo:2019tit,Figueroa:2019paj,Caldwell:2018giq,Bernal:2019lpc} and cosmic strings \cite{Cui:2017ufi,Cui:2018rwi,Chang:2019mza,Gouttenoire:2019kij,Gouttenoire:2019rtn} 
but has been mostly overlooked with regard to phase transitions (see \cite{Barenboim:2016mjm,Cai:2019cdl,Guo:2020grp,Ellis:2020nnr} for some recent studies in this direction).

\subsection{An intermediate period of matter domination} \label{sec:intermediate-md}

In this subsection we consider the effect of an intermediate period of matter domination (MD).  We will compare two scenarios.  
In the first scenario, GWs are produced at a temperature $T_\star$ and the universe is radiation dominated (RD) until CMB.  
In the second scenario, GWs are again produced at a temperature $T_\star$ but there is an intermediate period of matter domination between the temperatures $\Tstart$, where the universe goes from RD to MD, and $\Tend$, where it goes from MD to RD.  

\begin{figure}[t] \centering
\includegraphics[width=.7\textwidth]{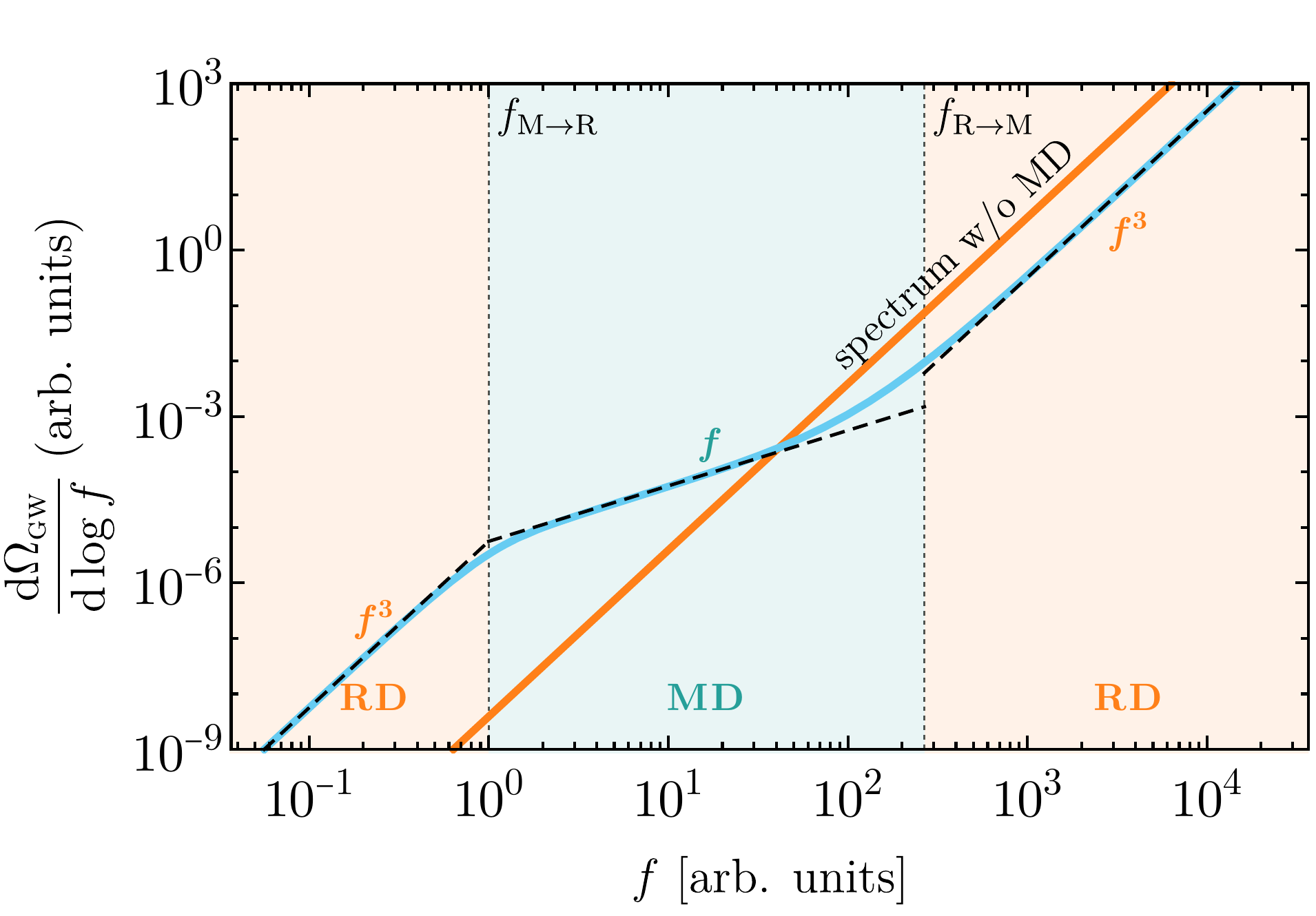}
\caption{The low-frequency spectrum of causality-limited GWs in the case with an intermediate period of matter domination (light blue) and the case with just radiation domination (orange).  The intermediate period of matter domination is present between the frequencies $\fmr$ to $\frm$.
The dashed black lines show the scaling as $f$ ($f^3$) during MD (RD).
High frequency modes have less power due to the presence of MD while low frequency modes have more power.
}
\label{fig: intermediate MD}
\end{figure}

A summary of the results in this section can be seen in Fig.~\ref{fig: intermediate MD} where we compare the numerical solutions for the spectrum in both cases.  We find that if one is interested in modes that enter the horizon before $\Tstart$, then the extra red-shifting induced by the period of MD simply decreases the amplitude of these modes.  
If one considers instead the modes that enter the horizon after $\Tend$ then their amplitude increases due to the red-shifting of the frequencies moving power from high frequencies to lower frequencies.
In what follows, we give an intuitive explanation of the results while in App.~\ref{app: analytical} we give a more analytical derivation.

Matter domination gives an entropy dump that increases the red-shift between the source and present day observers yielding two distinct effects on the spectrum.  The first is that the extra expansion dilutes the energy density, decreasing the overall power.  The second is that the frequencies themselves are red-shifted, an effect that increases the power at low frequency as it shifts the entire spectrum to lower frequencies.  The competition of these two effects determines the behavior of the spectrum.

\begin{figure}[h]\centering
\includegraphics[width=\textwidth]{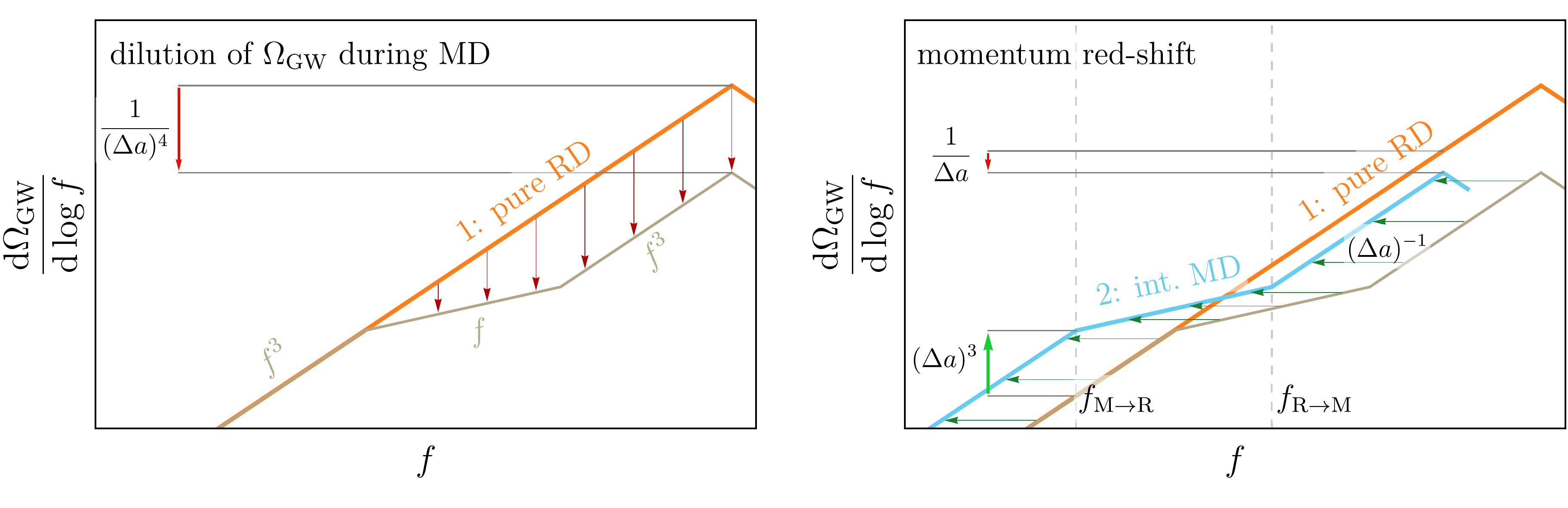}
\caption{A qualitative depiction of the two physical effects determining the shape of the causality-limited range of the GW spectrum in pure RD (scenario 1) and with an intermediate epoch of MD (scenario 2).
\textit{Left:} impact of the dilution of $\rho_\textsc{gw}$ with respect to the ambient energy density during MD for sub-horizon modes.
\textit{Right:} effect of the different redshift of the scale factor in the two scenarios.
See the text for more details.
}
\label{fig: expl int MD}
\end{figure}

The net effect of MD can be understood intuitively and is shown visually in Fig.~\ref{fig: expl int MD}.  
We first discuss the effect of the dilution of the energy density from the additional red-shifting.
For modes that enter the horizon before $\Tstart$, the overall power is redshifted by an additional $(\Delta a)^{-4}$ due to the additional expansion.  
Here we take 
\begin{equation}
\Delta a = \frac{(a_\star/a_0)_1}{(a_\star/a_0)_2} 
  = \left(\frac{\Tstart}{\Tend} \right )^{1/3} >1
\end{equation}
to be the difference between scenarios 1 (pure RD) and 2 (intermediate MD) in the amount of red-shifting between the phase transition and us. 
For modes that re-enter the horizon after $\Tend$, the modes were frozen out during matter domination and so their overall power is not modified.  
Intermediate frequencies interpolate between the two results with a $f$ scaling.
We now discuss the effect of the red-shifting of momenta.  
The observable GW frequency $f$ corresponds to the physical momentum $k/a$ of the GW, which in scenario 2 redshifts more by a factor $\Delta a$.
The $f^3$ fall off of the power spectrum during RD means that when comparing the two spectra at the same physical momenta, the power is enhanced by a factor of $(\Delta a)^3$ for all modes entering during RD. 
In totality, for modes that re-enter the horizon before $\Tstart$, these two effects combine to give a total suppression of $1/\Delta a$ while for modes that re-enter the horizon after $\Tend$, they combine to give a total enhancement of $(\Delta a)^3$.  Thus we find that an intermediate period of MD increases the power at low frequencies but suppresses it at high frequencies.

These estimates can be made more concrete as follows. 
In the first scenario where the universe was always radiation dominated, the spectrum of GWs is
\bea
\frac{\d \Omega_\textsc{gw,1}}{\d \log f} = \mathcal A_\textsc{gw,1} \frac{f^3}{f_{\star,1}^3} \qquad f_{\star,1} \sim \frac{T_\star T_0}{\Mp}
\eea
where $\mathcal A_\textsc{gw,1}$ is the amplitude of the GW spectrum at its peak and this scaling holds up to the peak frequency $f_{\star,1}$.  
Frequencies higher than this are sensitive to the properties of how the signal was generated.

As discussed before, there are two main effects of a period of intermediate matter domination.  
Firstly, the amplitude of the peak signal is decreased by $(\Delta a)^{-4}$, 
\bea
\mathcal A_\textsc{gw,2} = \mathcal A_\textsc{gw,1} \left ( \frac{\Tstart}{\Tend} \right )^{-4/3} .
\eea
Secondly, the frequencies are shifted by $1/\Delta a$ so that the new peak frequency is
\bea
f_{\star,2} = f_{\star,1} \left ( \frac{\Tstart}{\Tend} \right )^{-1/3}  \sim
\frac{T_\star T_0}{\Mp} \left ( \frac{\Tstart}{\Tend} \right )^{-1/3} .
\eea
Finally, the universe transitions to and from MD at the critical frequencies
\bea
\frm \sim \frac{\Tstart T_0}{\Mp} \left ( \frac{\Tstart}{\Tend} \right )^{-1/3}  \qquad 
\fmr \sim \frac{\Tend T_0}{\Mp}.
\eea 
From this starting point, the entire spectrum can be found by scaling as $f^3$ as long as the Universe is in RD and as $f^1$ as long as it is in MD.

The final spectrum is then
\begin{equation}
\frac{\d \Omega_\textsc{gw,2}}{\d \log f} =
\left\lbrace
\begin{aligned}
& \mathcal A_\textsc{gw,2} \frac{f^3}{f_{\star,2}^3} =  \frac{\d \Omega_\textsc{gw,1}}{\d \log f} \left ( \frac{\Tstart}{\Tend} \right )^{-1/3}   & f > \frm \\
& \mathcal A_\textsc{gw,2} \frac{\frm^3}{f_{\star,2}^3} \frac{f}{\frm} & \frm > f > \fmr \\
& \mathcal A_\textsc{gw,2} \frac{\frm^3}{f_{\star,2}^3} \frac{\fmr}{\frm}  \frac{f^3}{\fmr^3} =  \frac{\d\Omega_\textsc{gw,1}}{\d \log f} \left ( \frac{\Tstart}{\Tend} \right ) & f < \fmr
\end{aligned} \right.
\end{equation}
We see all of the effects that we had mentioned before.  At high frequencies, there is a suppression of $1/\Delta a = \left ( \Tstart/\Tend \right )^{-1/3}$ while at low frequencies there is an enhancement of $(\Delta a)^3 \sim \Tstart/\Tend$.

\begin{figure}[h] \centering
	\includegraphics[width=.6\textwidth]{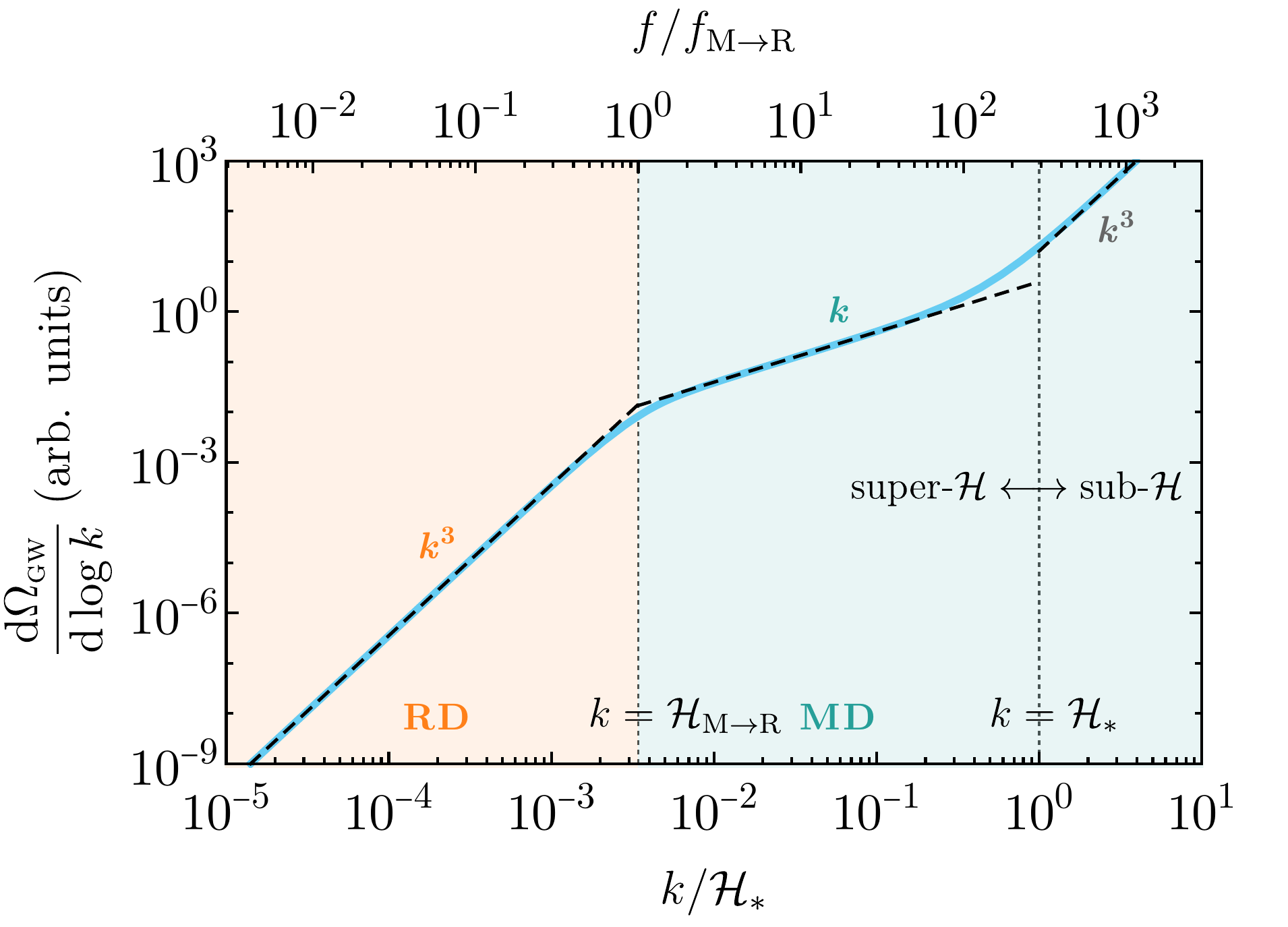}
	\caption{The spectrum of a GW produced during matter domination. Sub-Hubble causality enforces a $k^3$ scaling, while super-Hubble scaling gives $k$ during matter domination and $k^3$ during radiation domination.}
	\label{fig: PT during MD}
\end{figure}

A limit of the previous scenario is when the period of matter domination extends to the point in time when the GWs were produced. 
As shown in Fig.~\ref{fig: PT during MD}, in this case the power in modes that are sub-horizon when the spectrum is generated scales as $k^3$ while for those outside it goes as $k$.  
Thus even though this also has a transition from a $k^3$ scaling to a $k$ scaling, the origin for the $k^3$ scaling is different between the two scenarios, sub-horizon physics instead of radiation domination.
It is then an interesting question to see if it is possible to differentiate between the two.

As the transition from sub-horizon to MD and the one from RD to MD both lead to identical scaling away from the transition region, we cannot differentiate between the two based on their spectrum alone.  
However, the two transitions are not identical.  
Thus, in principle, if one sees the transition between a $k^3$ scaling to a $k^1$ scaling, one could use its shape to determine whether the $k^3$ scaling was due to radiation domination or modes being sub-horizon.

\begin{figure}[h] \centering
\includegraphics[width=.55\textwidth]{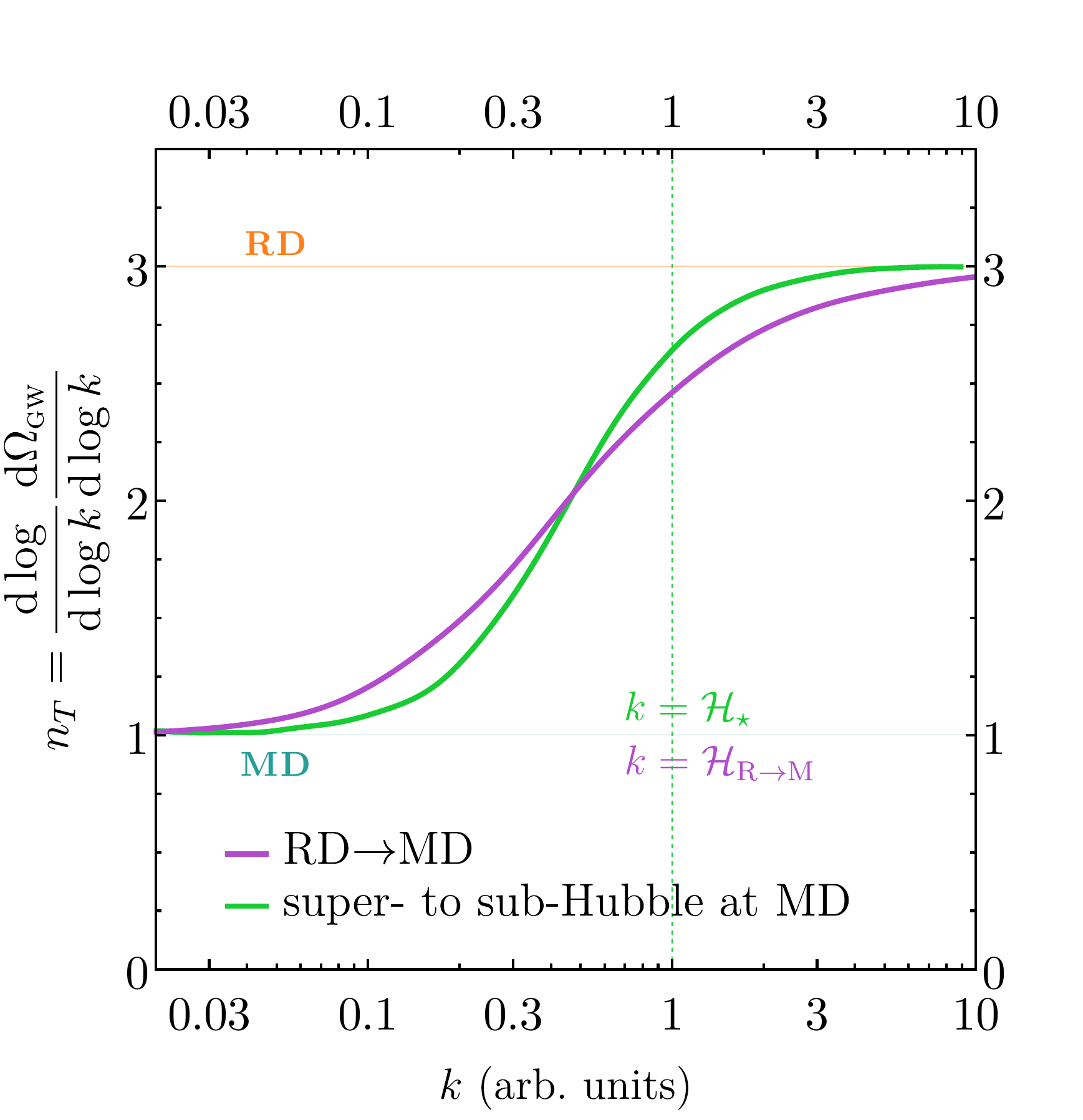} 
\caption{Comparison between the evolution of the tilt of GW spectra in two different scenarios: transition from \textit{sub}-horizon modes to MD super-horizon modes (green) and from RD \textit{super}-horizon modes to MD super-horizon modes (purple).
Sub-horizon and RD super-horizon modes both scale as $\d \OGW/\d \log k \sim k^3$ while MD super-horizon modes scale as $\d \OGW/\d \log k \sim k$ so that both scenarios have the exponent transition from $3$ to $1$.  
As is shown in the picture, the transition region is different, showing that the two scenarios can in principle be differentiated.}
\label{fig: compare nT (R vs sub) to M}
\end{figure}
We illustrate the difference between these two scenarios in Fig.~\ref{fig: compare nT (R vs sub) to M}.  
From this, one sees that the two different transitions are in fact distinguishable, but that their deviation is rather small. 
Without extra information from shorter wavelengths that are sensitive to the generation mechanism for the waves, one would need an extremely precise measurement of the transition region in order to determine which kind of transition it is.

\subsection{Using the GW spectrum to measure $w(t)$} \label{Sec: wt}

In the early universe, it is highly likely that $w$ was not constant and was a changing function of time.  In this section, we consider how to measure $w(t)$.  We will demonstrate that $w(t)$ can be to a very good approximation read off directly from the slope of $\d \OGW/\d \log k$. 

In an expanding universe with a constant $w$, the slope of $\d \OGW/\d \log k$ for super-horizon modes is given by Eq.~\eqref{Eq: superhorizonk} and repeated below for convenience
\bea
\frac{\d \OGW}{\d \log k} \sim k^{\frac{1+15 w}{1+3 w}} \, .
\eea
When one considers a generic $w(\tau)$, there is no longer an exact solution.  However, the intuition developed before still applies.  As mentioned before, we can approximate the GW as having a suppressed production coupled with being frozen out until $k = \H$.  These combined to give Eq.~\eqref{Eq: super-horizon}, repeated below for the sake of convenience
\bea
h =  \frac{a(\tk)}{a(\tau)} \frac{\Js}{\Hs} \sin  k \tau  \,. 
\eea
As long as $w'(\tau) \ll \H$, then we can proceed as before and obtain
\bea
\frac{\d \OGW}{\d \log k} \sim k^{\frac{1+15 w(\tau)}{1+3 w(\tau)}} \, ,
\eea
where, for each momentum $k$, $w(\tau)$ should be taken at the time $\tau$ when $k$ re-enters the horizon, $k=\H(\tau)$.
We see that in this approximation, one can simply read off $w(\tau)$ straight from the slope of $\d \OGW/\d \log k$.  
More explicitly
\bea
\frac{\d \log \left( \d \OGW/\d \log k \right)}{\d \log k} = \frac{1+15 w(\tau)}{1+3 w(\tau)} \, .
\eea
\begin{figure}[h] \centering
\includegraphics[width=.49\textwidth]{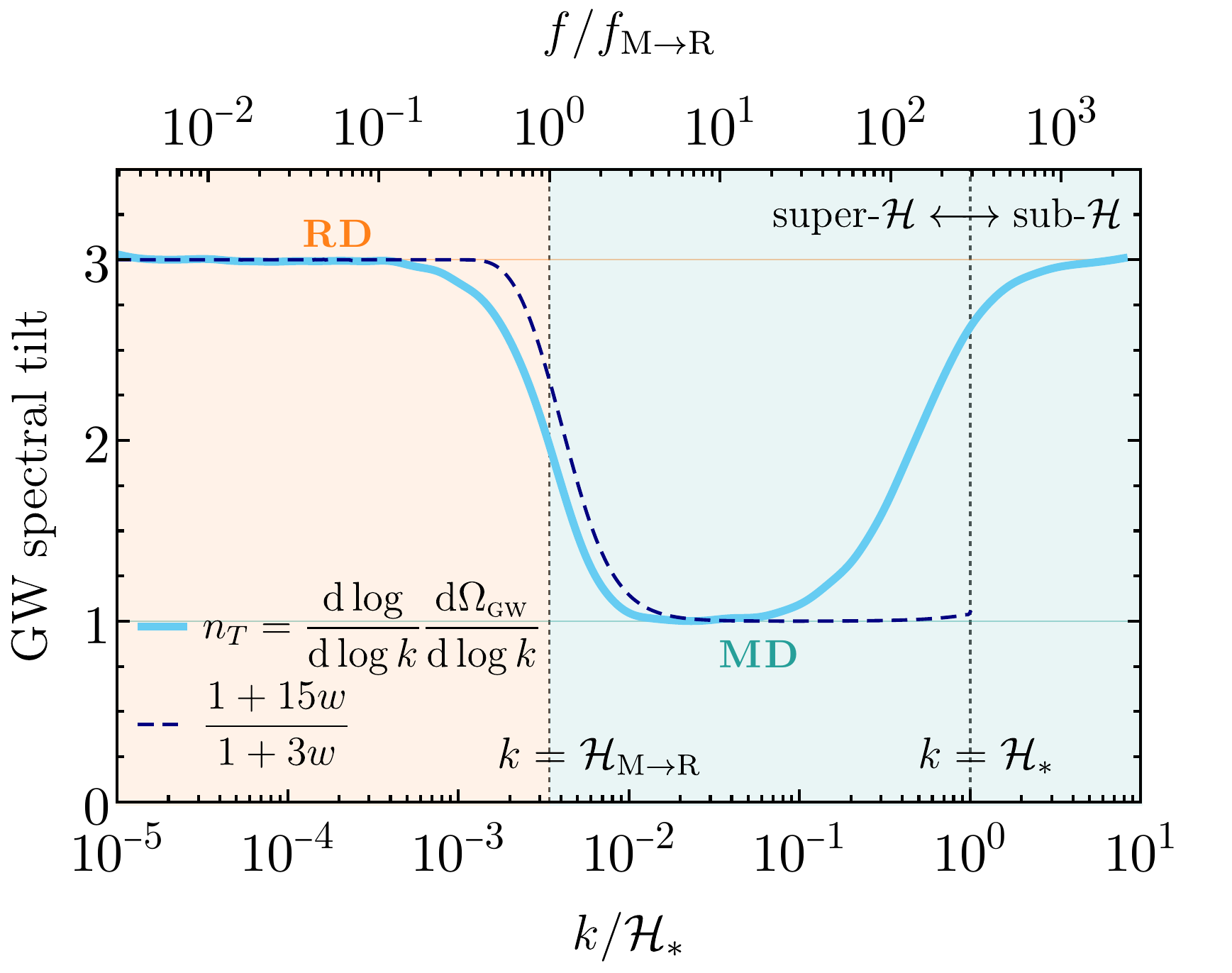} \hfill 
\includegraphics[width=.49\textwidth]{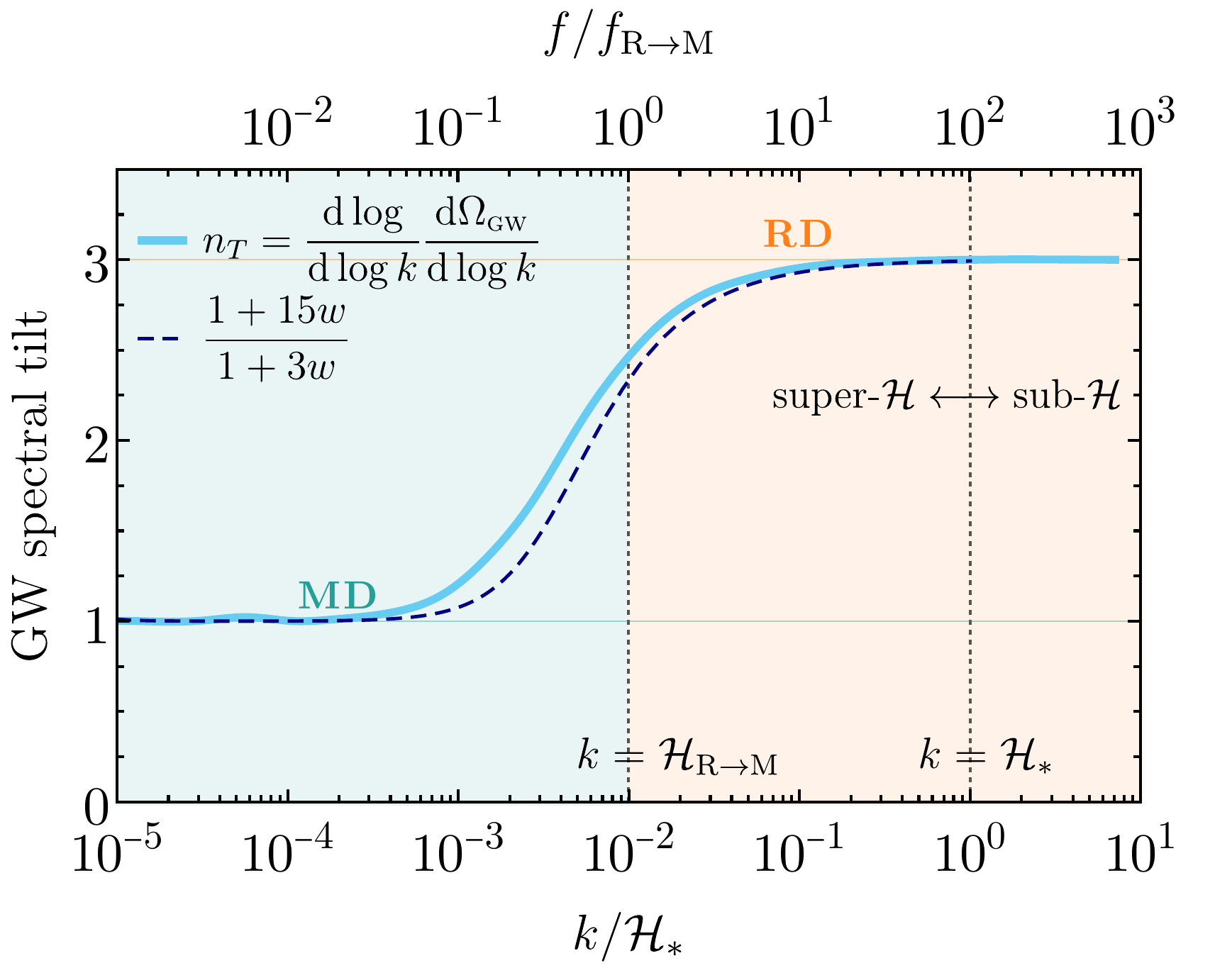}
\caption{A comparison between the exact equation of state of the universe versus what is measured from the slope of the GW spectrum.  In the solid blue line, we have the exact numerical result while in the dashed blue line, we have the approximation as measured from the slope of the GW spectrum.}
\label{fig: nT num an}
\end{figure}
\newline
Following this expression, one could in principle extract $w(\tau)$ by just measuring the tilt of the low-frequency tail of the GW spectrum.
The approximations made to obtain this simple expression are not valid during the transition region between matter and radiation domination.  
To see if they work in practice, we compared this procedure to several $w(\tau)$ numerically.  
The results are shown in Fig.~\ref{fig: nT num an}.  We see that this procedure is surprisingly accurate.

\section{Conclusion}
\label{Sec: conclusion}

In this paper, we have presented a physical picture of how to understand causality-limited gravitational waves.  
This picture allows one to easily estimate and predict the shape and behavior of low frequency gravity waves.  
As an example, we showed that generally sub-horizon and super-horizon modes behave very differently, allowing one to make model independent measurements of the conformal Hubble rate at which the gravity waves were created.  
Additionally, we showed how an intermediate period of MD has the effect of suppressing power at high frequencies and enhancing power at low frequencies.  

Perhaps the most surprising result is that we showed that free-streaming particles change the scaling of the GW power spectrum.  
Unlike the usual case where particles start free-streaming after GWs are produced, if there are free-streaming particles present when GWs are produced, they lead to a suppression in the production of super-horizon gravitational waves, bringing the scaling from $k^3$ down to a maximum of $k^4$.  
In the limit of a large number of free-streaming particles, oscillatory features are present demonstrating that interesting effects unrelated to causality can sculpt the spectrum of gravity waves.

There is much still left to be done with causality-limited gravitational waves.  
This range of the spectrum is special because its shape can be calculated from first principles.  
However, because causality sends the power to zero at low frequencies, a dedicated analysis would be needed to see to what extent gravity wave detectors such as LISA would be able to detect changes on this falling spectrum.
On a separate note, since preheating and reheating generated GWs are typically followed by a period of matter domination, they are more visible than previously expected.  
It would be interesting to see if these models naturally generate signals detectable by existing or future gravitational wave detectors.

\acknowledgments{
We thank Dani Figueroa and Raman Sundrum for stimulating discussions. 
We also thank Géraldine Servant and Peera Simakachorn for comments on the draft.
The authors also thank the KITP institute (From Inflation to the Hot Big Bang) where  part  of  this  work  was conducted, and the support of the NSF under the Grant No.\ PHY-1748958.
This research was supported in part  by  the  NSF  under Grants No.\ PHY-1914480, PHY-1914731 and by the Maryland Center for Fundamental Physics (MCFP). 
Research at Perimeter Institute is supported in part by the Government of Canada through the Department of Innovation, Science and Economic Development Canada and by the Province of Ontario through the Ministry of Colleges and Universities.
}

\appendix

\section{Analytical description of the GW spectrum with an intermediate period of matter domination} \label{app: analytical}
We start by using the physical intuition described in Sec.~\ref{Sec: intuition} to find an approximate analytical form of the GW spectrum in cosmologies with a non-standard expansion history between generation and observation of the GWs. 
We can approximate Eq.~\eqref{Eq: master} for super-horizon modes by
\begin{equation}
\begin{aligned}
& h'' + \frac{2 n_\star}{\tau} h'  = 0 \\
& h(\ts)  = 0 \, , \qquad  h'(\ts) = \Js \, ,
\end{aligned}
\end{equation}
where $n_\star = 2/(1+3w_\star)$ is fixed by the equation of state at $\ts$. As long as $n_\star>1/2$, these initial conditions quickly evolve to
\begin{equation}
h(\tau) = \frac{\Js}{\left(2 - 1/n_\star \right)\Hs} \, .
\end{equation}
Afterwards, $h$ remains constant while $k \ll \H$, in analogy with Eq.~\eqref{eq: initial intuition}, even if the equation of state (EOS) of the Universe changes.

Assuming a constant EOS during horizon entry, $k\tau \sim 1$, we must solve
\begin{equation}
\begin{aligned}
& h'' + \frac{2 n_k}{\tau} h' + k^2 h = 0 \, , \\
& h(\tau_i) = \frac{\Js}{\left(2 - 1/n_\star \right)\Hs} \, , \qquad h'(\tau_i) = 0 \, ,
\end{aligned}
\end{equation}
where the initial time, $\tau_i \ll 1/k$,  is chosen such that the equation of state is constant between $\tau_i$ and horizon entry, and $n_k$ is fixed by the equation of state of the Universe when the mode enters the horizon.
The solution to this equation in the $\tau_i \ll 1/k$ limit is given by
\begin{equation}
h(\tau) = \frac{2 \Js }{\left(2 - 1/n_\star \right)\Hs} \frac{\Gamma(n_k+1/2)}{\sqrt{\pi}} \left(\frac{2}{k \tau}\right)^{n_k-1} j_{n_k-1}(k \tau) \, .
\label{eq:solution-kdependent-eos}
\end{equation}
Taking the $k \tau \gg 1$ limit of the equation above we arrive at
\begin{equation}
h(\tau) \approx \frac{\Js }{\left(2 - 1/n_\star \right)\Hs} \frac{\Gamma(n_k+1/2)}{\sqrt{\pi}} \left(\frac{2}{k \tau}\right)^{n_k} \cos(k \tau - n_k \pi/2) \,.
\end{equation}
Once $k \gg \H$ the amplitude falls as $h \propto a^{-1}$, and hence, focusing only on the amplitude of $h$ we have
\begin{equation}
|h| \approx \frac{\Js }{\left(2 - 1/n_\star \right)\Hs} \frac{\Gamma(n_k+1/2)}{\sqrt{\pi}} \left(\frac{2}{k \tau_r}\right)^{n_k} \frac{a(\tau_r)}{a} \, ,
\label{eq:general-amplitude}
\end{equation}
where $\tau_r$ is some fixed reference time at which the EOS was still determined by $n_k$. Fixing the reference time $\tau_r$ allows us to compare different modes that enter horizon during an era of fixed $n$.

We will now use the result in Eq.~\eqref{eq:general-amplitude} to compare two scenarios to show the effects of non-standard expansion histories.  As before in the first scenario, GWs are produced at a temperature $T_\star$ at a scale factor $a_\star$ and the universe is radiation dominated until the CMB. 
In the second scenario, GWs are produced at the same temperature as before, $T_\star$, but there is an intermediate period of MD between the temperatures $\Tstart$ (where the universe becomes matter dominated) and $\Tend$ (where the universe becomes radiation dominated). 
For simplicity we will take the scale factor at production to be the same $a_\star$, which implies that the scale factor today will differ due to the difference in expansion history.

As discussed in Section~\ref{sec:intermediate-md}, one of the major effects of an intermediate period of MD is that the frequency spectrum of GWs is shifted by the additional expansion due to the matter dominated phase. Since we normalized the scale factor at generation $a_\star$ to be the same in both scenarios, the relation between the scale factor and plasma temperature is identical in both cases up until the temperature reaches $\Tstart$, when matter domination starts in the second scenario. The expansion history of both scenarios differ from that point on until the temperature gets to $\Tend$, after that point the scale factors corresponding to the same temperature (and therefore the scale factors at observation) are related by
\begin{equation}
a_2 = a_1 \left( \frac{\Tstart}{\Tend} \right)^{1/3} \, .
\end{equation}

Following  Eq.~\eqref{eq:general-amplitude}, the amplitude for modes that have entered the horizon in the first scenario is
\begin{equation}
h \approx \frac{\Js}{\Hs} \left( \frac{a_{\mtor,1} H_\mtor}{k} \right) \frac{a_{\mtor,1}}{a} \, .
\end{equation}
Therefore, the differential energy density is given by
\begin{equation}
\frac{d\rho_1}{d\log k} = \frac{H_\mtor^2}{2 (2 \pi)^3 G} \left( \frac{\Js}{\Hs} \right)^2 \left( \frac{a_{\mtor,1}}{a} \right)^4 k^3 \, .
\end{equation}

For the second scenario we need to separate between the modes that entered the horizon before, during and after the period of matter domination. The amplitude for modes that entered the horizon after the end of matter domination ($k < H_\mtor a_{\mtor,2}$) can be immediately found from Eq.~\eqref{eq:general-amplitude},
\begin{equation}
h \approx \frac{\Js}{\Hs} \left( \frac{a_{\mtor,2} H_\mtor}{k} \right) \frac{a_{\mtor,2}}{a} \, .
\end{equation}
Modes that entered the horizon during matter domination have a different form due to the difference in EOS and their amplitude is given by
\begin{equation}
h \approx \frac{\Js}{\Hs} \left( \frac{a_{\mtor,2} H_\mtor}{k} \right)^2 \frac{a_{\mtor,2}}{a} \, .
\end{equation}
Finally, modes that entered the horizon before matter domination are given by
\begin{equation}
h \approx \frac{\Js}{\Hs} \left(\frac{a_{\rtom} H_\rtom}{k}\right) \frac{a_\rtom}{a} \approx \left( \frac{a_{\rtom}^2 \Tstart^2}{a_{\mtor,2}^2 \Tend^2} \right) \frac{\Js}{\Hs} \left( \frac{a_{\mtor,2} H_\mtor}{k} \right) \frac{a_{\mtor,2}}{a} \, . 
\end{equation}
From which we find the differential energy density
\begin{multline}
\frac{d\rho_2}{d\log k} = \frac{H_\mtor^2}{2 (2 \pi)^3 G} \left( \frac{\Js}{\Hs} \right)^2 \left( \frac{a_{\mtor,2}}{a}\right)^4 k^3 
\, \times \\ \times
 \left\{ \begin{array}{cl}
1 \ , & \ k < a_{\mtor, 2} H_\mtor \\
\left( \frac{a_{\mtor,2} H_\mtor}{k} \right)^2 \ , & \ a_{\mtor, 2} H_\mtor < k < a_\rtom H_\rtom \\
\left( \frac{a_{\rtom}^2 \Tstart^2}{a_{\mtor,2}^2 \Tend^2} \right)^2 \ , & \ k > a_\rtom H_\rtom
\end{array}
\right. \, .
\end{multline}
Comparing this to the differential energy density for the RD only scenario, we can immediately see the $(\Delta a)^4$ dilution from the extra expansion discussed in the main text. Now, this distribution is in terms of conformal momentum $k$, which, due to the difference in scale factor normalization at late times correspond to distinct frequencies in the two cases. In order to compare the spectra of the two scenarios and their relation to experimental searches, it is more useful to compute the differential spectrum in terms of frequency, $f = k/(2 \pi a)$:
\begin{equation}
\frac{d \rho_1}{d\log f}(f) = \frac{H_\mtor^2}{2 G} \left( \frac{\Js}{\Hs} \right)^2 \left( \frac{a_{\mtor,1}}{a} \right)^4 a_1^3 f^3 \, .
\end{equation}
Using the fact that scale factors after the end of MD satisfy $a_{\mtor, 1}/a_1 = a_{\mtor,2}/a_2$, we find
\begin{equation}
\frac{d\rho_2}{d\log f}(f) = \left( \frac{a_{\mtor,2}}{a_{\mtor,1}} \right)^3 \frac{d \rho_1}{d\log f}(f)  \left\{ 
\begin{array}{cl}
1 \ , & \ f < f_\mtor \\
\left( \frac{f_\mtor}{f} \right)^2 \ , & \ f_\mtor < f < f_\rtom \\
\left( \frac{a_{\mtor,2}}{a_{\mtor,1}} \right)^{-4} \ , & \ f > f_\rtom
\end{array}
\right. \, ,	
\end{equation}
where $f_\rtom = a_{\rtom} H_\rtom/(2 \pi a_2) $ and $f_\mtor = a_{\mtor,2} H_\mtor/(2 \pi a_2)$, which reproduces the result discussed in the main text.

\bibliographystyle{JHEP}
\bibliography{Tilt-SGWB_refs.bib}

\end{document}